\documentclass[pre,tighten]{revtex4}

\usepackage{amssymb,amsmath}

\usepackage{graphicx}

\begin{document}
\title{Segregation by thermal diffusion of an intruder in a moderately dense granular fluid}
\author{Vicente Garz\'o}
\email{vicenteg@unex.es} \homepage{http://www.unex.es/eweb/fisteor/vicente/} \affiliation{Departamento de
F\'{\i}sica, Universidad de Extremadura, E-06071 Badajoz, Spain}

\begin{abstract}

A solution of the inelastic Enskog equation that goes beyond the weak dissipation limit and applies for moderate densities is used to determine the thermal diffusion factor of an intruder immersed in a dense granular gas under gravity.
This factor provides a segregation criterion that shows the transition between the Brazil-nut effect (BNE) and the reverse Brazil-nut effect (RBNE) by varying the parameters of the system (masses, sizes, density and coefficients of restitution).  The form of the phase-diagrams for the BNE/RBNE transition depends
sensitively on the value of gravity relative to the thermal gradient, so that it is
possible to switch between both states for given values of the parameters of the system. Two specific limits are considered with detail: (i) absence of gravity, and (ii) homogeneous temperature. In the latter case, after some approximations, our results are consistent with previous theoretical results derived from the Enskog equation. Our results also indicate that the influence of dissipation on thermal diffusion is more important in the absence of gravity than in the opposite limit. The present analysis extends previous theoretical results
derived in the dilute limit case [V. Garz\'o, Europhys. Lett. {\bf 75}, 521 (2006)] and is consistent with the findings of some recent experimental results.

{\bf Shortened title}: Segregation in dense granular gases

\end{abstract}

\draft 

\date{\today}
\maketitle

\section{Introduction}
\label{sec1}

One of the most important phenomena occurring in granular flows containing more than one species (a polydisperse system) is the segregation and mixing of dissimilar grains. This phenomenon, in which a homogeneous mixture of different species becomes spatially nonuniform by sorting themselves in terms of their masses and/or sizes, is of central interest in the field of granular matter mainly due to is industrial importance (powder metallurgy, pharmaceutical pills, glass and paint industries,$\ldots$). The resulting non-uniformity is usually an undesirable  property, although there are some applications in which one wants to force species segregation (e.g., the separation of mined ores). Unfortunately, in spite of its practical relevance, the physical mechanisms that govern mixing and separation processes are not well understood yet. This fact motivates the development of fundamental theories that predict accurately the bulk behavior of these systems in order to be able to control such processes.

It is well known that when a binary mixture constituted by one large ball and a number of smaller ones is subjected to vertical shaking in a container, usually the large particle (intruder) tends to climb to the top of the sample against gravity. This phenomenon is known as the Brazil-nut effect (BNE) and is one of the most puzzling problems of granular materials research \cite{RSPS87,KJN93,DRC93,CWHB96}. On the other hand, a series of experimental works \cite{SM98,HQL01} have also observed the reverse buoyancy effect, namely, under certain conditions the intruder can also sink to the bottom of the granular bed. This effect is known as the reverse Brazil-nut effect (RBNE). Several mechanisms have been proposed to explain the transition BNE/RBNE, for example, percolation \cite{RSPS87}, arching \cite{DRC93}, convection \cite{KJN93,CWHB96,LCBRD94}, inertia \cite{SM98}, condensation \cite{HQL01}, and  interstitial-fluid effects \cite{MLNJ01}. Among the different competing mechanisms, thermal diffusion becomes the most relevant one at large shaking amplitude where the sample of grains resembles a granular gas. In this regime, binary collisions prevail and kinetic theory can be a quite useful tool to analyze granular systems.

The thermal diffusion factor has been recently \cite{G06} evaluated for a {\em dilute} granular binary mixture
from a solution of the inelastic Boltzmann equation that applies for arbitrary degree of dissipation and takes into account the non-equipartition of energy. The results show that the relative position of the large particles (intruders of mass $m_0$) with respect to the small particles (gas particles of mass $m$) is determined by the sign of the control parameter $(m_0T-m T_0)/m_0T$, where $T_0$ and $T$ denote the temperatures of intruder and gas particles, respectively. While in a molecular or ordinary gas mixture this sign is fixed only by the mass ratio of the particles (since $T_0=T$), for a granular gas mixture it also depends on the temperature ratio because of the lack of equipartition. This segregation criterion compares well with molecular dynamics (MD) simulations carried out in the tracer or intruder limit case \cite{BRM05}. The objective here is to extend the analysis made
in Ref.\ \cite{G06} to higher densities by considering the revised Enskog theory. By extending the
Boltzmann analysis to high densities comparisons with MD simulations become practical and allow one to
quantitatively test the use of a hydrodynamic description for segregation in granular vibrated
mixtures. In addition, at higher densities, it is possible that other segregation mechanisms different from the
one identified in the dilute limit become relevant at those densities.

Previous theoretical attempts based on kinetic theory to describe segregation in {\em dense} granular mixtures
have also been reported. Thus, Jenkins and Yoon \cite{JY02} have developed a hydrodynamic theory for the
segregation of {\em elastic} particles, finding a criterion for segregation relatively close to the numerical
results obtained by Hong {\em et al.} \cite{HQL01}. Given that the criterion obtained in Refs.\ \cite{HQL01}
and  \cite{JY02} only applies for elastic particles, more recently Trujillo {\em et al.} \cite{TAH03} have
derived an evolution equation for the relative velocity of the intruder by using the kinetic theory proposed by
Jenkins and Mancini \cite{JM87} that only applies for {\em nearly} elastic particles. Interestingly, they considered the influence of the non-equipartition of granular energy (which is a generic feature of granular mixtures) on segregation through constitutive relations for the partial pressures.  However, the results reported by Trujillo {\em et al.} \cite{TAH03} have been derived by
neglecting the presence of temperature gradients in the bulk region so that, the segregation dynamics of
intruders is only driven by the gravitational force. Therefore, it appears that a complete theoretical description for the dynamics of BNE/RBNE in dense gases is still lacking.

As said before, the goal of this paper is to analyze the segregation by thermal diffusion of an intruder in a dense granular gas. The segregation criterion is obtained from a recent solution \cite{GDH07,GHD07} of the inelastic Enskog equation that covers some of the aspects not taken into account in previous works for dense systems \cite{JY02,TAH03} and extends previous results derived for dilute binary mixtures \cite{BRM05,G06} to higher densities. Specifically, (i) it takes into account the nonlinear dependence of the transport coefficients on dissipation so that the theory is expected to be applicable for a wide range of values of the coefficients of restitution, (ii) it considers the combined effect of gravity and thermal gradients on thermal diffusion and (iii) it applies for moderate densities. Consequently, the theory subsumes all previous analysis for both dense and dilute gases, which are recovered in the appropriate limits. In addition, the theoretical predictions are in qualitative agreement with some MD simulations \cite{BRM05,SUKSS06,GDH05} and are also consistent with previous experimental works \cite{BEKR03}. A preliminary report of some of the results presented here has been given in Ref.\ \cite{G08}.

The plan of the paper is as follows. First, the thermal diffusion factor $\Lambda$ is defined and evaluated in Sect.\ \ref{sec2} by using a hydrodynamic description. This factor provides a convenient measure of the separation or segregation of species generated by a thermal gradient in a multicomponent system. Once $\Lambda$ is expressed in terms of the pressure and the transport coefficients associated with the mass flux of impurities, these coefficients are explicitly determined in Sect.\ \ref{sec3} by solving the Enskog-Lorentz kinetic equation by means of the Chapman-Enskog method. This allows us to get $\Lambda$ as a function of the parameter space of the problem, namely, the mass and diameter ratios, the coefficients of restitution for collisions among gas-gas and intruder-gas particles and the solid volume fraction. In Sect.\ \ref{sec4}, the form of the phase-diagrams BNE/RBNE is widely investigated by varying the different parameters of the system. Moreover, a close comparison with the theoretical results \cite{JY02,TAH03} derived for thermalized dense gases is also carried out, showing that even in this limit the segregation criterion derived in this paper is more general than the one previously obtained since it covers the complete range of the parameter space of the system. The paper is closed in Sect.\ \ref{sec5} with a brief discussion of the results obtained in this paper.

\section{Hydrodynamic description for segregation by thermal diffusion}
\label{sec2}

We consider a moderately dense granular gas of inelastic hard disks ($d=2$) or spheres ($d=3$) of mass $m$ and diameter $\sigma$. The gas is in the presence of the gravitational field ${\bf g}=-g \hat{{\bf e}}_z$, where $g$ is a positive constant and $\hat{{\bf e}}_z$ is the unit vector in the positive direction of the $z$ axis. The particles collide with a constant coefficient of normal restitution $\alpha$. In the hydrodynamic description, it is assumed that the state of the gas is characterized by the local number density $n({\bf r},t)$, flow velocity ${\bf U}({\bf r},t)$, and temperature $T({\bf r},t)$. The time evolution of these fields is given by the balance hydrodynamic equations
\begin{equation}
\label{1} D_{t}n+n\nabla \cdot {\bf U}=0\;,
\end{equation}
\begin{equation}
\label{2} D_{t}{\bf U}+(mn)^{-1}\nabla \cdot {\sf P}={\bf g}\;,
\end{equation}
\begin{equation}
\label{3} D_{t}T+\frac{2}{dn}\left(\nabla \cdot {\bf q}+P_{ij}\nabla_j U_{i}\right) =-\zeta T\;,
\end{equation}
where $D_{t}=\partial _{t}+{\bf u}\cdot \nabla$ is the material time derivative. In the above equations, ${\sf P}$ is the pressure tensor, ${\bf q}$ is the heat flux and $\zeta$ is the cooling rate associated with the energy dissipation in collisions. The macroscopic balance equations (\ref{1})--(\ref{3}) are not entirely expressed in terms of the hydrodynamic fields, and thus do not comprise a closed set of equations. To close these equations one has to express the cooling rate and the fluxes as functionals of the fields. Such expressions are called ``constitutive relations'' and provide the link between the balance equations and a closed set of equations for the hydrodynamic fields alone. Such a closed set of equations defines {\em hydrodynamics} in its most general sense.

Let us assume now that some impurities of mass $m_0$ and diameter $\sigma_0>\sigma$ are added to the gas. Given that the impurities are present in tracer concentration, the problem is formally equivalent to study an intruder in a dense granular gas. This will be the terminology used in this paper. Collisions among intruder-gas particles are also {\em inelastic} and are characterized by the coefficient of normal restitution $\alpha_0$. It is also assumed that the presence of the intruder does not perturb the state of the gas and so, the flow velocity and temperature for the binary mixture composed by the dense gas plus the intruder are the same as those for the gas alone. Since the intruder may freely exchange momentum and energy with the gas particles, only the number density $n_0({\bf r},t)$ of the intruder is conserved. This continuity equation is given by
\begin{equation}
\label{4} D_{t}n_0+n_0\nabla \cdot {\bf U}+\frac{\nabla \cdot {\bf j}_0}{m_0}=0\;,
\end{equation}
where ${\bf j}_0$ is the mass flux for the intruder, relative to the local flow ${\bf U}$.

In this paper, we are interested in analyzing segregation by thermal diffusion of the intruder in a granular dense gas \cite{KCM87}.  Thermal diffusion is caused by the relative motion of the components of a mixture due to the presence of a thermal gradient. As a consequence of this motion, a steady state is reached in which the separating effect arising from the thermal diffusion is balanced by the remixing effect of ordinary diffusion \cite{KCM87}. From an experimental point of view, the amount of segregation parallel to the thermal gradient can be characterized by the thermal diffusion factor $\Lambda$. Phenomenologically, it is defined at the steady state in the absence of convection (zero flow velocity) through the relation
\begin{equation}
\label{5}
-\Lambda\frac{\partial \ln T}{\partial z} =\frac{\partial}{\partial z}\ln \left(\frac{n_0}{n}\right),
\end{equation}
where gradients only along the vertical direction ($z$ axis) have been assumed for simplicity. Let us assume that gravity and thermal gradient point in parallel directions (i.e., the bottom plate is hotter than the top plate, $\partial_z \ln T<0$). Thus, when $\Lambda >0$, the intruder tends to rise with respect to the fluid particles (i.e., $\partial_z\ln (n_0/n)>0$) while when $\Lambda <0$, the intruder falls with respect to the fluid particles (i.e., $\partial_z\ln (n_0/n)<0$). The former situation is referred to as the Brazil-nut effect (BNE) while the latter is called the reverse Brazil-nut effect (RBNE).

As said before, we consider an inhomogeneous non-convecting steady state with only gradients in the $z$ direction. Since ${\bf U}={\bf 0}$, then the mass flux ${\bf j}_0$ vanishes in the steady state according to the balance equation (\ref{4}). Moreover, clearly the pressure tensor is diagonal for this state and so, $P_{ij}=p\delta_{ij}$ where $p$ is the hydrostatic pressure. In this case, the momentum balance equation (\ref{2}) reduces to
\begin{equation}
\label{6} \frac{\partial p}{\partial z}=-\rho g,
\end{equation}
where $\rho=mn$ is the mass density of the gas particles. As will be shown later, the spatial dependence of the pressure $p$ is through its dependence on the number density $n$ and the temperature $T$. As a consequence, Eq.\ ({\ref{6}) can be written more explicitly as
\begin{equation}
\label{7} \frac{p}{T}\partial_zT+\frac{\partial p}{\partial
n}\partial_z n=-\rho g,
\end{equation}
where the partial derivative $\partial_n p$ will be computed once the equation of state for the gas is obtained. To close the problem of determining $\Lambda$ one needs a constitutive equation for the mass flux. Symmetry considerations yield
\begin{equation}
\label{8} j_{0,z}=-\frac{m_0^2}{\rho}D_{0}
\frac{\partial n_0}{\partial z}-\frac{m_0 m}{\rho}D\frac{\partial n}{\partial z}-
\frac{\rho}{T}D^T\frac{\partial T}{\partial z},
\end{equation}
where $D_{0}$ is the so-called kinetic diffusion coefficient, $D$ is the mutual diffusion coefficient and $D^T$ is the thermal diffusion coefficient. These transport coefficients measure the contribution of each independent gradient to the mass flux of intruders. The condition $j_{0,z}=0$ (applied to (\ref{8})) along with the momentum equation (\ref{7}) allows one to express the thermal diffusion factor $\Lambda$ defined by Eq.\ (\ref{5}) in terms of the thermal gradient $\partial_zT$ and gravity. Its expression is
\begin{equation}
\label{9} \Lambda=\frac{\beta D^{T*}-(p^*+g^*)(D_{0}^*+D^*)}{\beta D_{0}^*}.
\end{equation}
Here, $\beta=p^*+n\partial_n p^*$, $p^*=p/nT$, and we have introduced the reduced transport coefficients
\begin{equation}
\label{10}
D^{T*}=\frac{\rho \tau}{n_0T}D^T,\quad D_0^{*}=\frac{m_0^2 \tau}{\rho T}D_0,\quad D^{*}=\frac{m_0 \tau}{n_0 T}D,
\end{equation}
where $\tau$ is a collision frequency (to be chosen later).  In addition,
\begin{equation}
\label{11}
g^*=\frac{\rho g}{n\left(\frac{\partial T}{\partial z}\right)}<0
\end{equation}
is a dimensionless parameter measuring the gravity relative to the thermal gradient.
This quantity measures the competition between these two mechanisms ($g$ and
$\partial_z T$) on segregation.

To get the dependence of the thermal diffusion factor on the parameters of the system, the explicit form of the transport coefficients and the equation of state is needed. This can be achieved by solving the inelastic Enskog equation by means of the Chapman-Enskog method. This will be carried out in the next Section.

\section{Transport coefficients}
\label{sec3}

We adopt now a kinetic theory point of view and start from the Enskog kinetic equation for the system. Thus, all the macroscopic (or hydrodynamic) properties of interest of the system (dense gas plus intruder) are determined from the one-particle velocity distribution functions of the gas  $f({\bf r}, {\bf v},t)$ and the intruder $f_0({\bf r}, {\bf v},t)$. Since thermal diffusion is given in terms of the transport coefficients $D_0$, $D$, and $D^T$ associated with the mass flux of the intruder, our goal here is to solve the corresponding Enskog-Lorentz equation for the intruder by applying the Chapman-Enskog method \cite{CC70} to first order in the spatial gradients. The Enskog equation neglects velocity correlations among particles which are about to collide, but it takes into account the dominant spatial corrections to the Boltzmann equation (which only applies for dilute gases) due to excluded volume effects. Although the first assumption (molecular chaos hypothesis) can be questionable at high densities, there is substantial evidence in the literature of the accuracy of the Enskog theory for densities outside the dilute limit (moderate densities) and
values of dissipation beyond the quasielastic limit. As a matter of fact, this is the only available theory at present for making explicit calculations of the transport properties of moderately dense gases.

In order to fluidize the system in most of the experiments energy is added to the gas by the bottom wall which vibrates in a given way. Due to this external injection of energy, the system reaches a steady state whose properties far away from the boundaries (bulk domain) are expected to be insensitive to the details of the driving forces. However, due to the mathematical complexities associated with the use of vibrating boundary conditions, here we will
introduce a stochastic external force, coupling the velocity of each particle to a white noise (stochastic thermostat). This kind of forcing, which has been shown to be relevant for some two-dimensional experimental configurations with a rough vibrating piston \cite{PEU02}, has been used by many authors \cite{thermostat} in the past years to analyze different problems, such as segregation in granular binary mixtures \cite{TAH03,G08}. Although the relationship of these external forces with real vibrating walls is not clear to date, some results \cite{G06} derived in driven steady states for the temperature ratio by using the stochastic driving method agree quite well with molecular dynamics simulations \cite{SUKSS06} of shaken mixtures. This agreement suggests that this driving method can be seen as a plausible approximation for comparison with experiments. In addition, the advantage of such a driving mechanism is that it lends itself to theoretical progress. Under the above conditions, the Enskog-Lorentz equation for the intruder reads
\begin{equation}
\label{12} \left(\frac{\partial}{\partial t}+{\bf v}\cdot \nabla -\frac{1}{2}
\frac{\zeta T}{m}\frac{\partial^2}{\partial v^2}+{\bf g}\cdot
\frac{\partial}{\partial {\bf v}}\right) f_0({\bf r}, {\bf v},t)=J_{0}[{\bf v}|f_0(t),f(t)],
\end{equation}
where the collision operator $J_{0}[{\bf v}|f_0(t),f(t)]$ is
\begin{eqnarray}
\label{13} J_{0}[{\bf r}_1, {\bf v}_1|f_0(t),f(t)]&=&\overline{\sigma}^{d-1} \int d\mathbf{v}_{2}\int
d\widehat{{\boldsymbol\sigma}}\Theta ( \widehat{{\boldsymbol\sigma}}\cdot
\mathbf{g})(\widehat{{\boldsymbol\sigma}}
\cdot \mathbf{g})  \notag \\
&&\times \left[ \alpha_0^{-2}\chi_0\left( \mathbf{r}_{1},\mathbf{r} _{1}-\overline{{\boldsymbol\sigma}}\right) f_0(\mathbf{r} _{1},\mathbf{v}_{1}^{\prime \prime
};t)f(\mathbf{r}_{1}-\overline{{\boldsymbol\sigma}}
,\mathbf{v}_{2}^{\prime \prime };t)\right.  \notag \\
&&\left. -\chi_0\left( \mathbf{r}_{1},\mathbf{r}_{1}+\overline{{\boldsymbol\sigma}}
\right) f_0(\mathbf{r}_{1},\mathbf{v}
_{1};t)f(\mathbf{r}_{1}+\overline{{\boldsymbol\sigma}},\mathbf{v}_{2};t)\right] .
\end{eqnarray}
Here, $\overline{{\boldsymbol\sigma}}=\overline{\sigma}\widehat{{\boldsymbol\sigma}}$, $\overline{\sigma}=(\sigma_0+\sigma)/2$, $\widehat{{\boldsymbol\sigma}}$ is a unit vector along the centers of the two colliding spheres, $\alpha_0$ ($0\leq \alpha_0 \leq 1$) is the coefficient of
restitution for intruder-gas collisions, and $\chi_0$ is the pair correlation function for intruder-gas
pairs at contact. The precollisional velocities are given by
\begin{equation}
{\bf v}_{1}^{\prime \prime}={\bf v}_{1}-\frac{m}{m_0+m}\left( 1+\alpha _{0}^{-1}\right)
(\widehat{{\boldsymbol {\sigma }}}\cdot {\bf
g})\widehat{{\boldsymbol {\sigma }}} ,\nonumber\\
\end{equation}
\begin{equation}
 {\bf v}_{2}^{\prime \prime}={\bf v}_{2}+\frac{m_0}{m_0+m}\left( 1+\alpha
_{0}^{-1}\right) (\widehat{{\boldsymbol {\sigma }}}\cdot {\bf g})\widehat{ \boldsymbol
{\sigma}}. \label{14}
\end{equation}
Upon writing Eq.\ (\ref{12}) we have assumed that the system is driven by means of a stochastic Langevin force representing Gaussian white noise \cite{WM96}. This force is written as $\boldsymbol{\mathcal {F}}_0=m_0{\boldsymbol \xi}$, where the covariance of the stochastic acceleration has been chosen to be the same for the gas particles and the intruder \cite{HBB00,BT02}. In the context of the Enskog equation (\ref{12}), this external force is represented by a Fokker-Planck collision operator of the form $-\frac{1}{2}(\zeta T/m) \partial^2/\partial v^2$. Note that the covariance of the external force has been taken to achieve a constant  temperature in the homogeneous state. The generalization of the force to the {\em inhomogeneous} case is essentially a matter of choice and here, for simplicity, we have assumed that the stochastic force has the same form as in the homogeneous case except that now $\zeta$ and $T$ are in general functions of space and time. This simple choice has been widely used in ordinary gases to analyze nonlinear transport in shearing systems \cite{GS03}.

As said before, the main goal of this Section is to compute the mass flux ${\bf j}_0$ to first order in the spatial gradients. It is defined as
\begin{equation}
{\bf j}_{0}=m_{0}\int d{\bf v}\,{\bf V}\,f_0({\bf r},{\bf v},t), \label{15}
\end{equation}
where ${\bf V}={\bf v}-{\bf U}$ is the peculiar velocity. At a kinetic level, another interesting quantity for the intruder is its local temperature defined as
\begin{equation}
\label{16.1} T_0({\bf r}, t)=\frac{m_0}{d n_0({\bf r}, t)}\int \; d{\bf v}\, V^2 f_0({\bf r},{\bf v},t).
\end{equation}
This quantity measures the mean kinetic energy of the intruder. As confirmed by computer simulations \cite{BRM05, SUKSS06,computer}, experiments \cite{exp} and kinetic theory calculations \cite{GD99}, the global temperature
$T$ and the temperature of the intruder $T_0$ are in general different.

In order to determine the mass flux, we solve the Enskog-Lorentz equation by means of the Chapman-Enskog (CE) expansion \cite{CC70}. This method assumes the existence of a {\em normal} solution in which all the space and time dependence of $f_0$ occurs through the hydrodynamic fields $n_0$, $n$, ${\bf u}$ and $T$. The CE
procedure generates the normal solution explicitly by means of an expansion in gradients of the fields:
\begin{equation}
\label{15.1} f_0=f_0^{(0)}+\epsilon f_0^{(1)}+\cdots,
\end{equation}
where $\epsilon$ is a formal parameter measuring the nonuniformity of the system. The
application of the CE method to the Enskog equation for polydisperse granular
mixtures has been carried out very recently \cite{GDH07,GHD07} in the undriven case. The extension of these
calculations to the driven case is straightforward. We only display here the final expressions for the transport coefficients with some technical details given in Appendix \ref{appA}. As for elastic collisions, the coefficients $D_0$, $D$, and $D^T$ are given in terms of the solutions of linear integral equations, which can be approximately solved by considering the leading terms in a Sonine polynomial expansion. Here, we have considered for simplicity the first Sonine approximation. In dimensionless form, the transport coefficients are defined by Eq.\ (\ref{10}) with $\tau=n\sigma^{d-1}\sqrt{2T/m}$ \cite{note}. Their explicit forms have been obtained in Appendix \ref{appA} with the result
\begin{equation}
\label{16} D_{0}^*=\frac{\gamma}{\nu_D^*},
\end{equation}
\begin{equation}
\label{17} D^{T*}=-\frac{M}{\nu_D^*}\left(p^*-\frac{\gamma}{M}\right)+
\frac{(1+\omega)^d}{2\nu_D^*}\frac{M}{1+M} \chi _{0}^{(0)}\phi(1+\alpha _{0}),
\end{equation}
\begin{equation}
\label{18} D^{*}=-\frac{M}{\nu_D^*}\beta+\frac{1}{2\nu_D^*}
\frac{\gamma+M}{1+M}\frac{\phi}{T}\left(\frac{\partial\mu_0}{\partial
\phi}\right)_{T,n_0}(1+\alpha_{0}).
\end{equation}
Here, $\gamma\equiv T_0/T$ is the temperature ratio, $M\equiv m_0/m$ is the mass ratio,
$\omega\equiv \sigma_0/\sigma$ is the size ratio, $\chi_0^{(0)}$ is the pair correlation function for intruder-gas evaluated at zeroth order,
\begin{equation}
\label{19} \nu_D^*=\frac{2\pi^{(d-1)/2}}{d\Gamma\left(\frac{d}{2}\right)}\left(\frac{\overline{\sigma}}{\sigma}
\right)^{d-1}\frac{\chi_0^{(0)}}{1+M}\sqrt{\frac{M+\gamma}{M}}(1+\alpha_0),
\end{equation}
and $\mu_0$ is the chemical potential of the intruder. When the granular gas is driven by means of a
stochastic thermostat, the temperature ratio $\gamma$ is determined from the
requirement \cite{BT02,DHGD02}
\begin{equation}
\label{20}
\gamma \zeta_0^*=M\zeta^*,
\end{equation}
where $\zeta_0^*$ is the (reduced) cooling rate associated with the partial temperature $T_0$. The expressions of the (reduced) cooling rates $\zeta^*$ and $\zeta_0^*$ in the local equilibrium approximation are given by \cite{GDH07}
\begin{equation}
\label{21}
\zeta^*\equiv \frac{\zeta^{(0)}}{\tau}=\frac{\sqrt{2}\pi^{(d-1)/2}}{d\Gamma(d/2)}\chi^{(0)} (1-\alpha^2),
\end{equation}
\begin{equation}
\label{22}
\zeta_0^*\equiv \frac{\zeta_0^{(0)}}{\tau}=\frac{4\pi^{(d-1)/2}}{d\Gamma\left(\frac{d}{2}\right)}\left(\frac{\overline{\sigma}}{\sigma}
\right)^{d-1}
\frac{\chi_0^{(0)}}{1+M}\sqrt{\frac{M+\gamma}{M}}(1+\alpha_0)\left[1-\frac{\gamma+M}{2\gamma(1+M)} (1+\alpha_{0})\right],
\end{equation}
where $\chi^{(0)}$ is the pair correlation function for the granular gas evaluated at zeroth-order. Moreover, the (reduced) pressure $p^*$ is \cite{GD99b,L05}
\begin{equation}
p^*=1+2^{d-2} \chi^{(0)} \phi(1+\alpha), \label{23}
\end{equation}
where
\begin{equation}
\phi\equiv \frac{\pi^{d/2}}{2^{d-1}d\Gamma(d/2)} n\sigma^d \label{24}
\end{equation}
is the solid volume fraction.

\begin{figure}
\includegraphics[width=0.5 \columnwidth,angle=0]{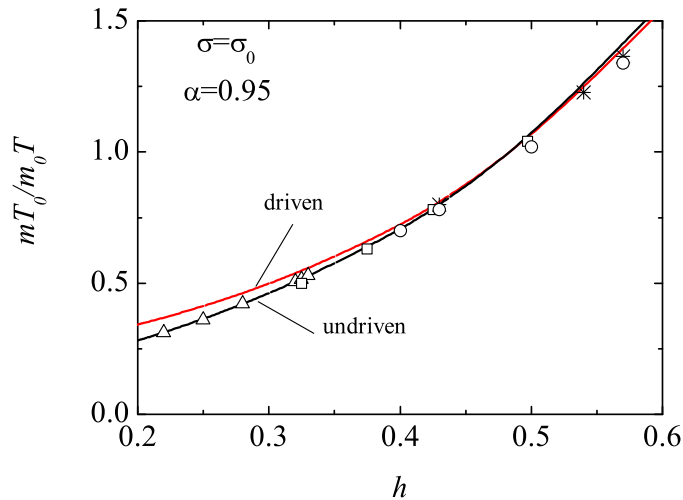}
\caption{(color online) Plot of the ratio of the mean square velocities $mT_0/m_0T$ as a function of $h\equiv m(1+\alpha_0)/2(m+m_0)$ for hard disks ($d=2$), $\alpha=0.95$, $\sigma_0/\sigma=1$ in the case of a dilute gas ($\phi=0$). The solid lines are the theoretical predictions given by Eqs.\ (\ref{20}) and (\ref{25}) and the symbols are MD simulation results obtained by Brey {\em et al.}  \cite{BRM05} for different values of the mass ratio: $m_0/m=2$ (triangles), 1 (squares), 0.75 (stars), and  0.5 (circles).  \label{fig1}}
\end{figure}
Before considering the dependence of the transport coefficients on dissipation, let us illustrate first the dependence of the temperature ratio $\gamma$ on the parameters of the problem. It must be noted that condition (\ref{20}) to determine the temperature ratio differs from the one derived in the undriven (free cooling) case \cite{GD99}, where $\gamma$ is obtained by requiring the equality of the cooling rates, i.e.,
\begin{equation}
\label{25}
\zeta^*=\zeta_0^*.
\end{equation}
\begin{figure}
\includegraphics[width=0.5 \columnwidth,angle=0]{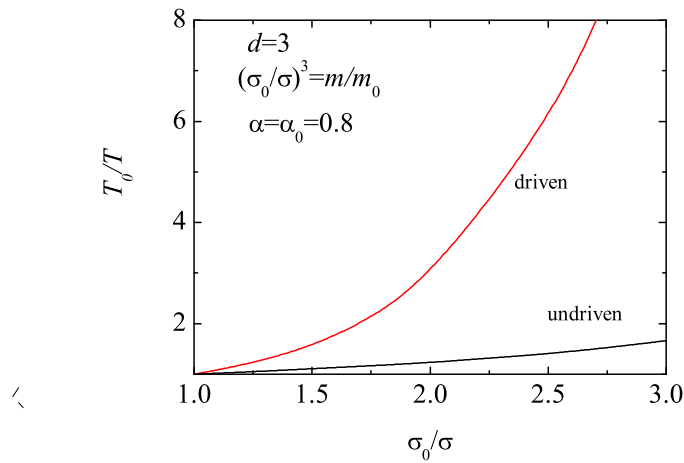}
\caption{(color online) Plot of the temperature ratio $T_0/T$ as a function of the size ratio $\sigma_0/\sigma$
for inelastic hard spheres ($d=3$) with $\phi=0.1$, $\alpha=\alpha_0=0.8$ and when the intruder and the gas particles have the same mass density [$
m_0/m=(\sigma_0/\sigma)^3$]. The lines are the kinetic theory results obtained in the driven and undriven cases.   \label{fig2}}
\end{figure}
Since the condition to determine the ratio $T_0/T$  is different in the driven and undriven states, it is interesting to explore the similarities and differences between the temperature ratios in both situations. Note that, according to Eq.\ (\ref{22}), the dependence of $\zeta_0^*$ on $\gamma$ is through the ratio of mean square velocities $\Theta\equiv mT_0/m_0T$. In terms of the parameter $\Theta$, the conditions (\ref{20}) and (\ref{25}) are cubic equations with a unique real, positive solution. In particular, the behavior of the solution in the limit $\Theta \to 0$ for undriven homogeneous states has been analyzed by Santos and Dufty \cite{SD01,SD02}, where a change similar to a second order phase transition has been shown. In Fig.\ \ref{fig1}, we plot $\Theta$ as a function of the dimensionless quantity $h\equiv m(1+\alpha_0)/2(m+m_0)$ for hard disks ($d=2$), $\alpha=0.95$, $\sigma_0/\sigma=1$ and in the case of a dilute gas ($\phi=0$ and so, $\chi^{(0)}=\chi_0^{(0)}=1$). The theoretical predictions obtained from the conditions (\ref{20}) and (\ref{25}) indicate that $\Theta$ is a function only of the parameter $h$ for given values of $\alpha$ and $\sigma_0/\sigma$ \cite{SD01}. Molecular dynamics simulation results obtained by Brey {\em et al.} \cite{BRM05} in an open vibrated granular gas for different values of the mass ratio have been also included. It is apparent that, for the range of values explored in Fig.\ \ref{fig1}, the heating mechanism slightly affects the value of $\Theta$ since the theoretical curves obtained from the driven and undriven conditions yield quite  identical results. Moreover, the agreement between theory (driven and undriven cases) and simulation data is very good, except perhaps for small values of $h$ where the results obtained in the undriven case compare with simulation data better than those derived in the driven case. On the other hand, significant discrepancies between the results obtained with and without a thermostat for the temperature ratio $T_0/T$ are observed in Fig.\ \ref{fig2}, where $T_0/T$ is plotted versus the size ratio $\sigma_0/\sigma$ for a moderately dense gas ($\phi=0.1$). Here,
the intruder and the gas particles are composed by spheres of the same material and therefore, the same mass density [i.e., $m_0/m=(\sigma_0/\sigma)^3$]. To evaluate the ratio $T_0/T$, one has to know the explicit forms of $\chi^{(0)}$ and $\chi_0^{(0)}$. In the case of hard spheres ($d=3$), a good approximation for
$\chi^{(0)}$ is provided by the Carnahan-Starling form \cite{CS69}
\begin{equation}
\label{n.1} \chi^{(0)}=\frac{1-\frac{1}{2}\phi}{(1-\phi)^3},
\end{equation}
while the intruder-gas pair correlation function is given by \cite{B70}
\begin{equation}
\label{n.2}
\chi_0^{(0)}=\frac{1}{1-\phi}+3\frac{\omega}{1+\omega}\frac{\phi}{(1-\phi)^2}+2
\frac{\omega^2}{(1+\omega)^2}\frac{\phi^2}{(1-\phi)^3}.
\end{equation}
It is apparent that the disagreement found in Fig.\ \ref{fig2} illustrates the fact that the heating mechanisms affect in general nonequipartition even in the bulk of the system \cite{WM08}. Although the lack of simulation data in Fig.\ \ref{fig2} prevent us to assess the reliability of both theoretical predictions, it must be noted that a recent comparison \cite{G06} between MD simulations of agitated binary mixtures \cite{SUKSS06} and kinetic theory  results based on the driven condition (\ref{20}) shows a good agreement in contrast with the predictions obtained from the undriven relation (\ref{25}) where significant discrepancies between the latter and simulation appear especially as the size ratio increases (see Fig.\ 1 of Ref.\ \cite{G06}). More simulation data are needed to make quantitative comparisons between theories based on homogeneously heated granular systems and boundary-driven problems in order to asses the reliability of the above theoretical predictions.
\begin{figure}
\includegraphics[width=0.5 \columnwidth,angle=0]{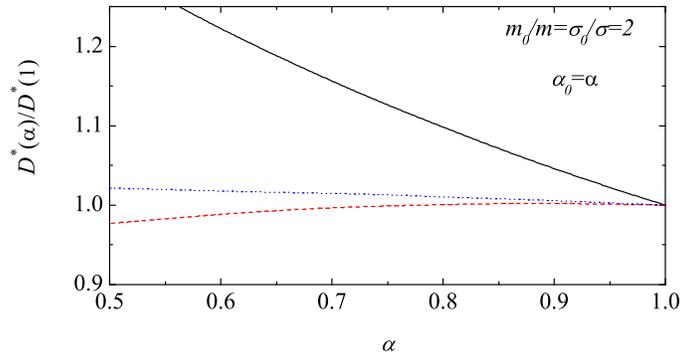}
\caption{(color online) Plot of the reduced mutual diffusion coefficient
$D^*(\alpha)/D^*(1)$ as a function of the (common) coefficient of restitution
$\alpha=\alpha_0$ for inelastic hard spheres ($d=3$) with $m_0/m=\sigma_0/\sigma=2$ and
three values of the solid volume fraction: $\phi=0$ (solid line), $\phi=0.2$ (dashed line) and $\phi=0.4$
(dotted line).
\label{fig3}}
\end{figure}
\begin{figure}
\includegraphics[width=0.5 \columnwidth,angle=0]{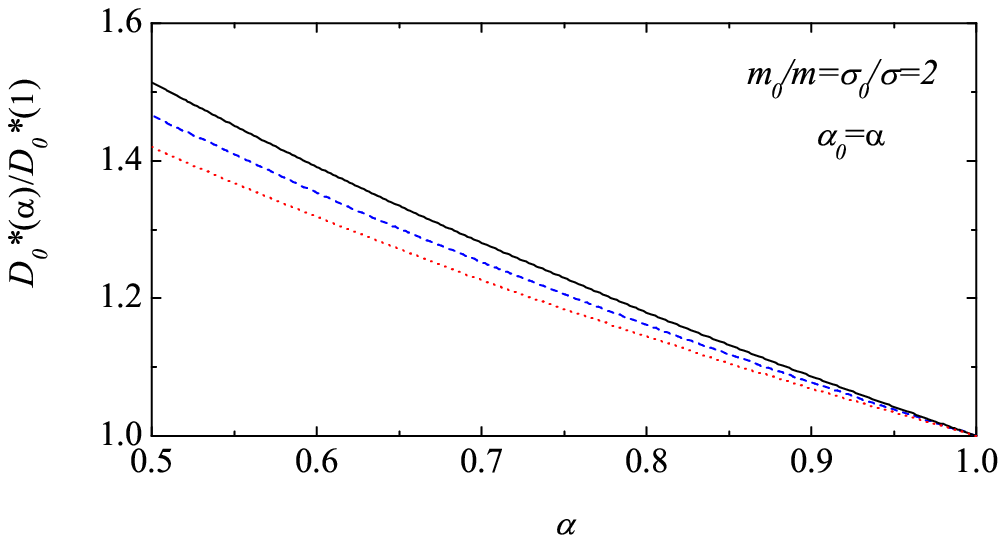}
\caption{(color online) Plot of the reduced kinetic diffusion coefficient
$D_0^*(\alpha)/D_0^*(1)$ as a function of the (common) coefficient of restitution
$\alpha=\alpha_0$ for inelastic hard spheres ($d=3$) with $m_0/m=\sigma_0/\sigma=2$ and
three values of the solid volume fraction: $\phi=0$ (solid line), $\phi=0.2$ (dashed line) and $\phi=0.4$
(dotted line).
\label{fig4}}
\end{figure}
\begin{figure}
\includegraphics[width=0.5 \columnwidth,angle=0]{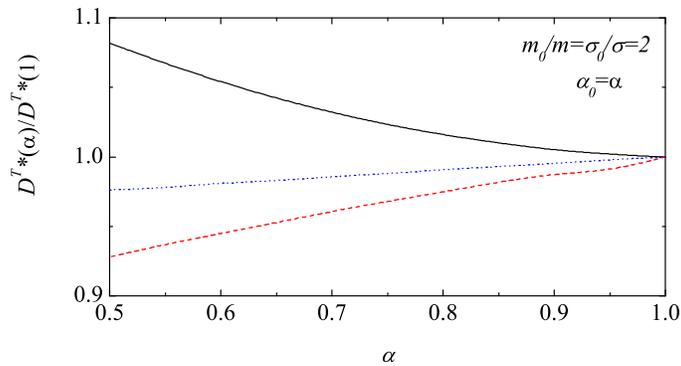}
\caption{(color online) Plot of the reduced thermal diffusion coefficient
$D^{T*}(\alpha)/D^{T*}(1)$ as a function of the (common) coefficient of restitution
$\alpha=\alpha_0$ for inelastic hard spheres ($d=3$) with $m_0/m=\sigma_0/\sigma=2$ and
three values of the solid volume fraction: $\phi=0$ (solid line), $\phi=0.2$ (dashed line) and $\phi=0.4$
(dotted line).
\label{fig5}}
\end{figure}

Now, the dependence of the transport coefficients on dissipation is considered.  According to Eq.\ (\ref{18}), in order to get the explicit dependence of $D_0^*$ on $\alpha$ one has to give the form of the chemical potential $\mu_0$. The expression for the chemical potential of the intruder consistent with the approximation (\ref{n.2}) is
\cite{RG73}
\begin{eqnarray}
\label{n.3} \frac{\mu_0}{T}&=&C_3+\ln n_0-\ln (1-\phi)+3\omega \frac{\phi}{1-\phi}+
3\omega^2\left[\ln (1-\phi)+\frac{\phi(2-\phi)}{(1-\phi)^2}\right]\nonumber\\
& & -\omega^3\left[2\ln (1-\phi)+\frac{\phi(1-6\phi+3\phi^2)}{(1-\phi)^3}\right],
\end{eqnarray}
where $C_3$ is a constant. Figures \ref{fig3}, \ref{fig4} and \ref{fig5} show the reduced coefficients $D^*(\alpha)/D^*(1)$, $D_0^*(\alpha)/D_0^*(1)$, and $D^{T*}(\alpha)/D^{T*}(1)$, respectively, versus the (common) coefficient of restitution $\alpha=\alpha_0$ for inelastic hard spheres with $m_0/m=\sigma_0/\sigma=2$ and several values of the solid volume fraction : $\phi=0$ (dilute gas), $\phi=0.2$ (moderately dense gas) and $\phi=0.4$ (dense gas). All the transport coefficients have been reduced with respect to their elastic values $D^*(1)$, $D_0^*(1)$ and  $D^{T*}(1)$. We observe that in general the deviation from the functional form for elastic collisions is significant, especially in the case of the kinetic diffusion coefficient $D_0^*$. This deviation becomes more important as the density of the gas decreases. Although not shown in the above figures, a comparison with the results obtained in the undriven case \cite{GF09} shows again quantitative differences between the results derived with and without a thermostat and so, as expected \cite{GM02,WM08}, the latter does not play a neutral role in the mass transport.

\begin{figure}
\begin{center}
\begin{tabular}{lr}
\resizebox{8cm}{!}{\includegraphics{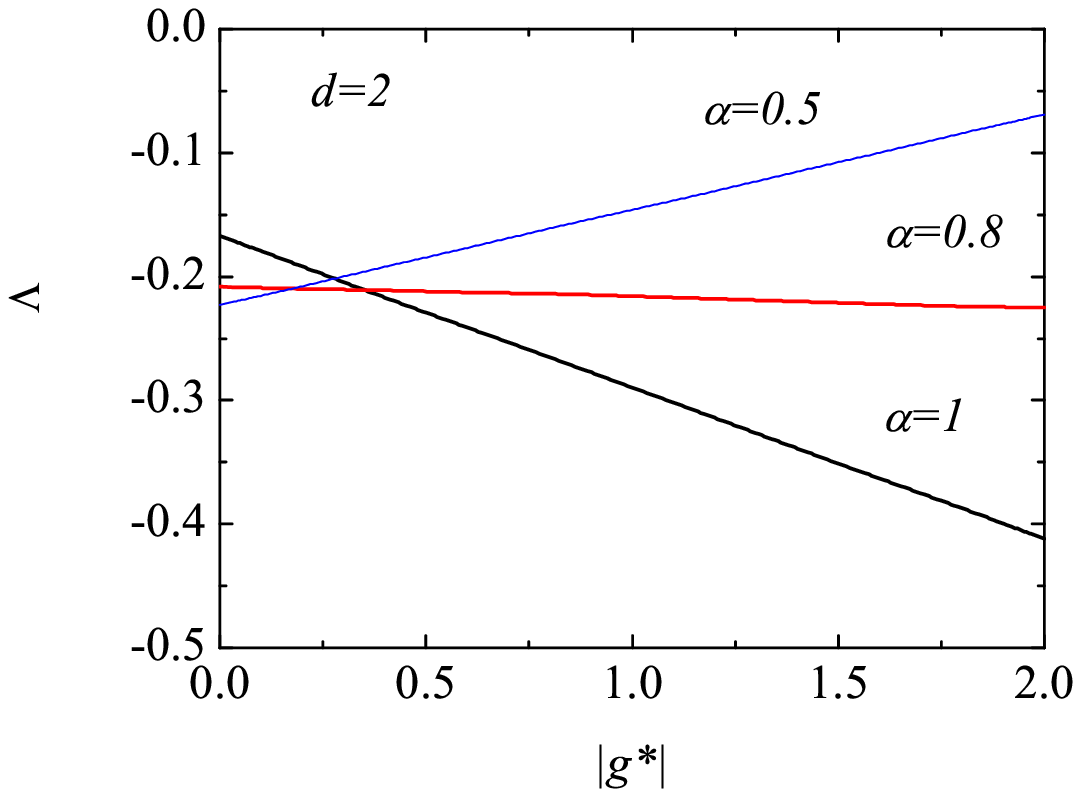}}&\resizebox{7.75cm}{!}{\includegraphics{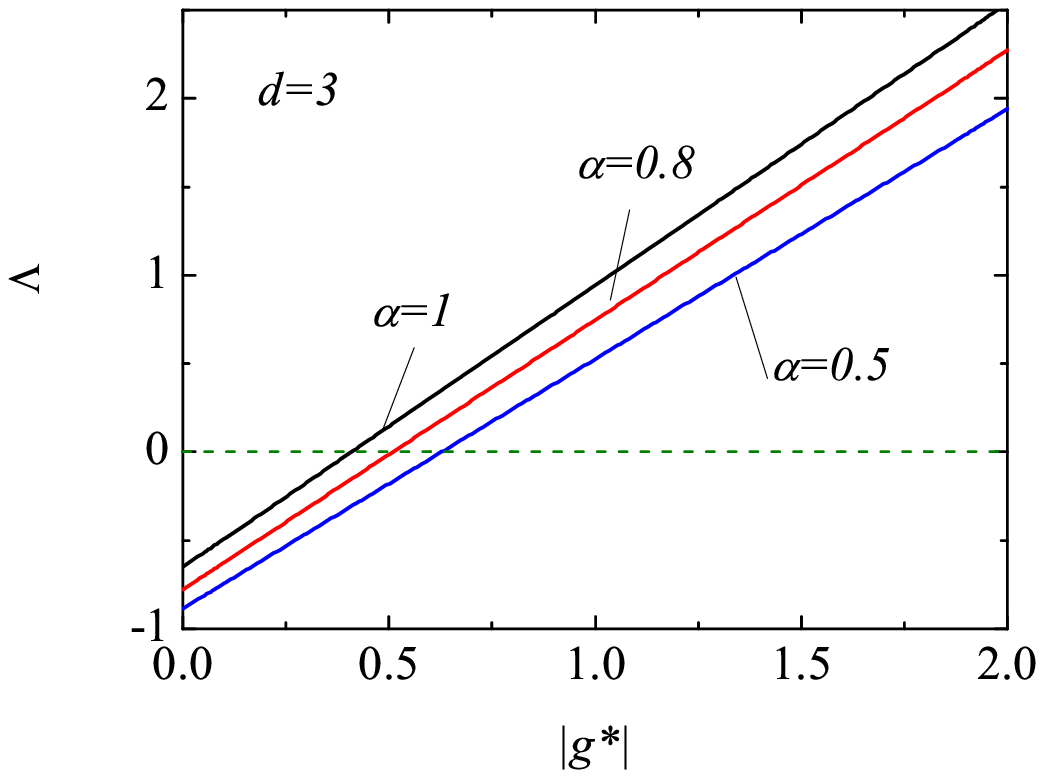}}
\end{tabular}
\end{center}
\caption{(color online) Plot of the thermal diffusion factor $\Lambda$ versus the (reduced) gravity $|g^*|$
for $m_0/m=\sigma_0/\sigma=2$, $\phi=0.2$ and three values of the (common) coefficient of restitution $\alpha=\alpha_0$.
The left panel is for hard disks $(d=2)$ while the right panel is for hard spheres $(d=3)$.  \label{fig6}}
\end{figure}

\section{Thermal diffusion factor. Phase diagrams for the BNE/RBNE transition}
\label{sec4}

Once the form of the transport coefficients is known, the thermal diffusion factor $\Lambda$ can be explicitly obtained when one substitutes Eqs.\ (\ref{16})--(\ref{18}) for $D_0^*$, $D^{T*}$ and $D^*$, respectively, and Eq.\ (\ref{23}) for $p^*$ into Eq.\ (\ref{9}). This gives the dependence of $\Lambda$ in terms of the parameter space of the problem. This space is sixfold: the dimensionless gravity $g^*$, the mass ratio $m_0/m$, the size ratio $\sigma_0/\sigma$, the coefficients of restitution $\alpha$ and $\alpha_0$ and the the solid volume fraction $\phi$. According to Eq.\ (\ref{9}), $\Lambda$ is a linear function of gravity $g^*$, as is illustrated in Fig.\ \ref{fig6} where thermal diffusion is plotted as a function of $|g^*|$ for hard disks ($d=2$) and spheres ($d=3$). We have considered the system $m_0/m=\sigma_0/\sigma=2$, $\phi=0.2$ and several values of the (common) coefficient of restitution $\alpha=\alpha_0$. In the three-dimensional case, $\chi^{(0)}$, $\chi_0^{(0)}$ and $\mu_0$ are given by Eqs. \ (\ref{n.1})--(\ref{n.3}), respectively, while for the two-dimensional case, $\chi^{(0)}$ and $\chi_0^{(0)}$ are approximately given by \cite{JM87}
\begin{equation}
\label{n.4} \chi^{(0)}=\frac{1-\frac{7}{16}\phi}{(1-\phi)^2},
\end{equation}
\begin{equation}
\label{n.5}
\chi_0^{(0)}=\frac{1}{1-\phi}+\frac{9}{8}\frac{\omega}{1+\omega}\frac{\phi}{(1-\phi)^2}.
\end{equation}
For hard disks, the form of the chemical potential consistent with the above approximations is \cite{S08}
\begin{equation}
\label{n.6}
\frac{\mu_0}{T}=C_2+\ln n_0-\ln (1-\phi)+\frac{1}{4}\omega \left[\frac{9\phi}{1-\phi}+\ln(1-\phi)\right]+
\frac{1}{8}\omega^2\left[
\frac{\phi(7+2\phi)}{(1-\phi)^2}-\ln(1-\phi)\right],
\end{equation}
where $C_2$ is a constant. It is apparent for the case analyzed here that the RBNE is dominant for hard disks in all the range of values of gravity (except for $\alpha=0.5$ where a change to BNE is expected for values of $|g^*|$ larger than 2) while the RBNE is only dominant at small $|g^*|$ for hard spheres. Thus, as already noted in Ref.\ \cite{G08}, for given values of $m_0/m$, $\sigma_0/\sigma$, $\alpha$, $\alpha_0$ and $\phi$ there is a critical value $|g_c^*|$ such that a transition BNE$\Leftrightarrow$ RBNE (or RBNE$\Leftrightarrow$ BNE) is observed for $|g^*|>|g_c^*|$. We see that the value of $|g_c^*|$ increases with dissipation.

The condition $\Lambda=0$ provides the segregation criterion for the transition BNE/RBNE. Given that the parameter $\beta=p^*+\phi \partial_\phi p^*$ and the kinetic diffusion coefficient $D_0^*=\gamma/\nu_D^*$ are both positive, then according to Eq.\ (\ref{9}), the line delineating the regimes between BNE and RBNE is obtained from the relation
\begin{equation}
\label{26}
\beta D^{T*}=(p^*+g^*)(D_{0}^*+D^*).
\end{equation}
This condition can be written more explicitly when one takes into account the explicit forms (\ref{16})--(\ref{18}) of the transport coefficients. The result is
\begin{eqnarray}
\label{27}
& & g^*(\gamma-M)+\phi\left[(\gamma-M g^*)\frac{\partial p^*}
{\partial \phi}-M\frac{p^*-1}{\phi}g^*\right]\nonumber\\
& &
+\frac{(1+\omega)^d}{2}\frac{M}{1+M}\chi
_{0}^{(0)}\phi(1+\alpha_{0})\left[\left(p^*+g^*\right)\frac{M+\gamma}{M}\Delta-\beta\right]=0,
\end{eqnarray}
where
\begin{equation}
\label{28}
\Delta\equiv \frac{(1+\omega)^{-d}}{\chi_{0}^{(0)}T}\left(\frac{\partial \mu_0}{\partial \phi}\right)_{T,n_0}.
\end{equation}
Equation (\ref{28}) contains all the information necessary to describe the segregation due to thermal (or Soret) diffusion of an intruder in a moderately dense granular fluid. The first term on the left hand side measures essentially the influence of the non-equipartition of the granular energy on segregation. This term vanishes in the absence of the gravitational force. The second and third terms are proportional to the solid volume fraction $\phi$ and so, they account for the effects of density on thermal diffusion. These latter two terms vanish in the dilute limit case ($\phi\to 0$). The influence of each term on the segregation criterion (\ref{27}) depends on the specific values of dissipation (which is for instance the main responsible for the energy non-equipartition), density, mechanical parameters of the system and/or (reduced) gravity.

Before exploring the dependence of the parameter space on the form of the phase diagrams BNE/RBNE, it is instructive to consider some special limit situations. Thus, when the intruder and particles of the gas are mechanically equivalent ($m_0=m$, $\sigma_0=\sigma$, and $\alpha_0=\alpha$), then as expected the two species do not segregate. This is consistent with Eq.\ (\ref{27}) [or Eq.\ (\ref{26})] since in this limit case $D^{T*}=D_0^*+D^*=0$
and so, $\Lambda=0$ for any value of $\alpha$ and $\phi$. On the other hand, in the case of a dilute gas ($\phi\to 0$), then Eq.\ (\ref{27}) yields
\begin{equation}
\label{29}g^*(\gamma-M)=0.
\end{equation}
In the absence of gravity, Eq.\ (\ref{29}) holds trivially so that, the intruder does not segregate in a dilute gas when $g^*=0$. This result is due to the failure of the first Sonine approximation to accurately describe this special situation since segregation would appear if one would retain higher order terms in the Sonine polynomial expansion \cite{KCM87,GF09}. On the other hand, when $|g^*|\neq 0$, the solution to Eq.\ (\ref{29}) is
\begin{equation}
\label{30} \frac{T_0}{T}=\frac{m_0}{m}.
\end{equation}
This result agrees with recent results derived from the Boltzmann equation \cite{BRM05,G06}. Note that, due to the lack of energy equipartition, the criterion (\ref{30}) is rather complicated since it involves all the parameter space. The segregation criterion (\ref{30}) compares well with MD simulation results for the case of the steady state of an open vibrated granular system in the absence of macroscopic fluxes \cite{BRM05}.

\begin{figure}
\begin{center}
\begin{tabular}{lr}
\resizebox{8cm}{!}{\includegraphics{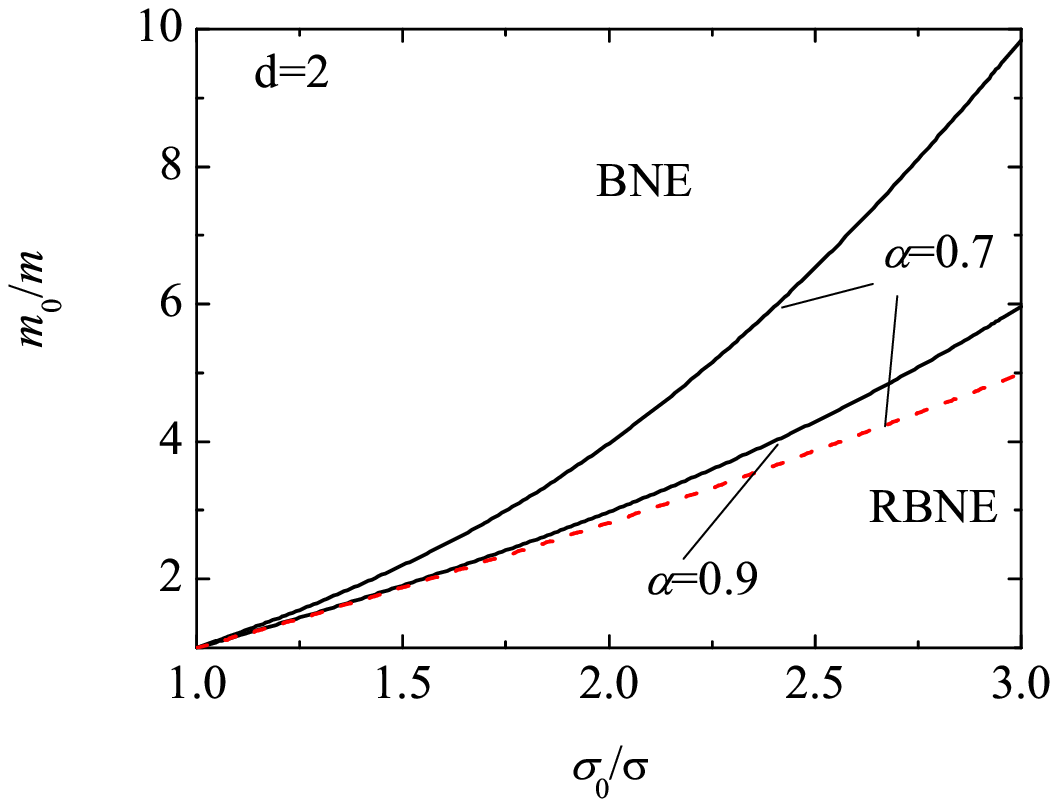}}&\resizebox{7.75cm}{!}{\includegraphics{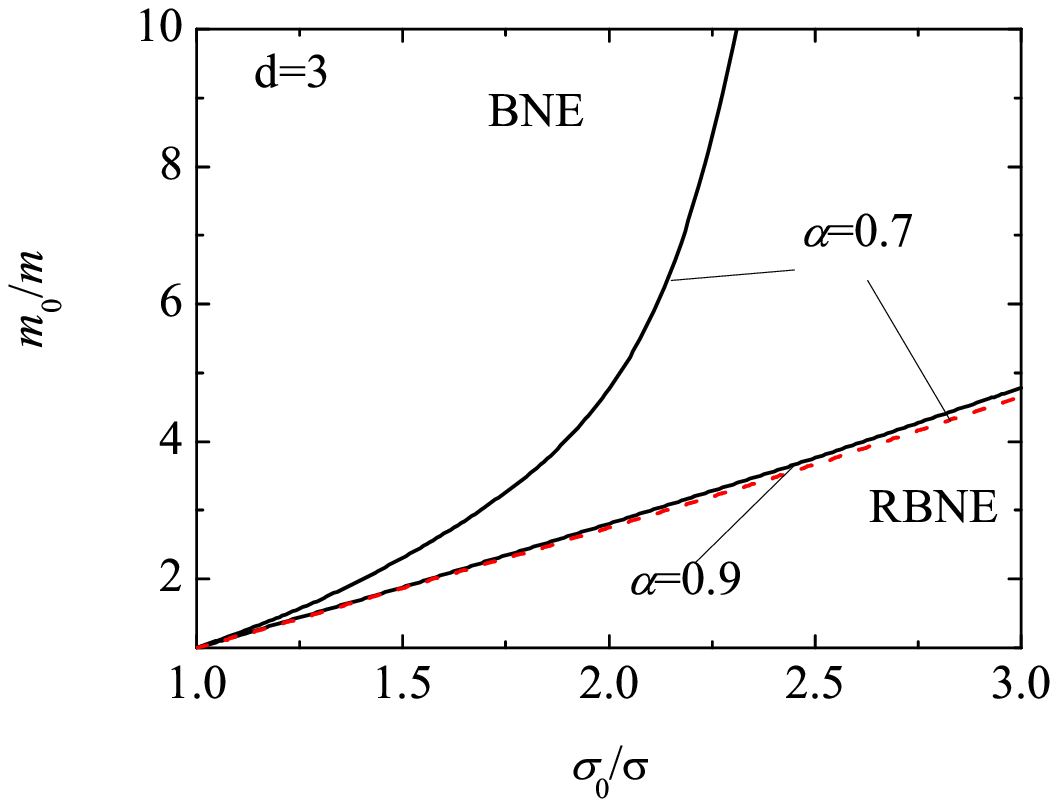}}
\end{tabular}
\end{center}
\caption{(color online) Phase diagram for BNE/RBNE for $\phi=0.25$ in the absence of gravity ($g^*=0$)
for two values of the (common) coefficient of restitution $\alpha=\alpha_0$. Points above the curve correspond to $\Lambda>0$ (BNE) while points below the curve correspond to $\Lambda<0$ (RBNE). The dashed line is the result obtained for $\alpha=0.7$ assuming energy equipartition ($T_0=T$).
The left panel is for hard disks $(d=2)$ while the right panel is for hard spheres $(d=3)$.  \label{fig7}}
\end{figure}

According to Eq.\ (\ref{27}), segregation is driven and sustained by both gravity and temperature gradients. The combined effect of both $g$ and $\partial_zT$ on thermal diffusion is through the dimensionless gravity $g^*<0$ defined by Eq.\ (\ref{11}). This parameter measures the competion between both mechanisms on segregation. To separate the influence of each one of the terms appearing in (\ref{27}) on segregation, we consider now some specific cases.
\begin{figure}
\includegraphics[width=0.5 \columnwidth,angle=0]{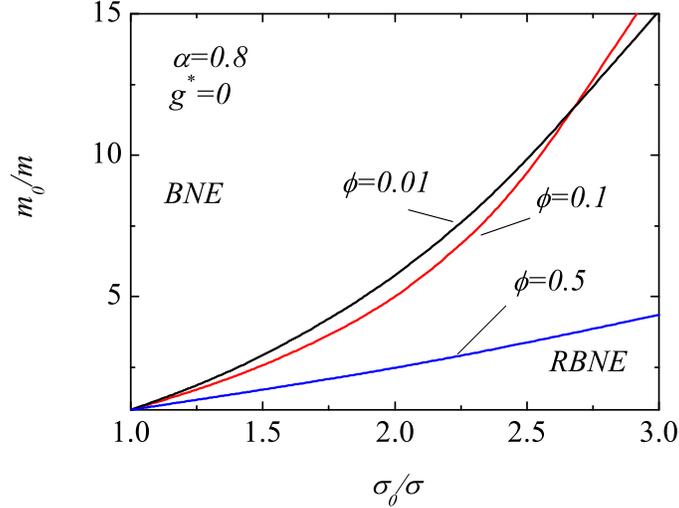}
\caption{(color online) Phase diagram for BNE/RBNE for inelastic hard spheres for $\alpha=\alpha_0=0.8$, $|g^*|=0$ and three different values of the solid volume fraction $\phi$.
\label{fig7bis}}
\end{figure}
\subsection{Absence of gravity ($|g^*|\to 0$)}

Lets us focus first on a system in which effects of the temperature gradient are assumed to dominate over gravity so that the latter can be neglected. This limit is usually considered in MD simulations (see, for example, the simulations carried out by Galvin {\em et al.} \cite{GDH05}). In this case, $|g^*| \to 0$ and Eq.\ (\ref{27}) reduces to
\begin{equation}
\label{31}
\gamma \phi \frac{\partial p^*}{\partial \phi}=\frac{(1+\omega)^d}{2}\frac{M}{1+M}\chi
_{0}^{(0)}\phi(1+\alpha_{0}) \left[\phi \frac{\partial p^*}{\partial \phi}+p^*\left(1-\Delta \frac{\gamma+M}{M}\right)\right].
\end{equation}
Of course, this equation is trivially satisfied in the case of a dilute gas ($\phi=0$). Beyond the dilute limit, the influence of each term in (\ref{31}) is still intricate. As an illustration, Figure \ref{fig7} shows the phase diagram in the $\{m_0/m, \sigma_0/\sigma\}$--plane at a total volume fraction of $\phi=0.25$ (moderately dense gas) and two different values of the (common) coefficient of restitution $\alpha_0=\alpha$. It is apparent that, in the absence of gravity, the main effect of dissipation is to reduce the size of the BNE. This effect is more significant in the case of hard spheres than in the case of disks. We observe that in general the RBNE is dominant for both small mass ratio and/or large size ratio. In order to assess the impact of the non-equipartition of granular energy on segregation, we have also plotted the corresponding phase diagram for $\alpha=0.7$ but assuming that $T_0=T$. The comparison between both curves clearly shows the significant influence of the temperature differences on thermal diffusion in the absence of gravity. This is consistent with the recent MD-findings of Galvin {\em et al.} \cite{GDH05} where they showed that non-equipartition driving forces for segregation are comparable to other driving forces for systems displaying moderate level of non-equipartition. Figure \ref{fig7bis} illustrates the influence of the volume fraction on the phase-diagram for a three dimensional system at moderate level of dissipation ($\alpha=0.8$). It is apparent that the role played by the density is quite important since the range of size and mass ratios for which the RBNE exists increases with decreasing $\phi$.

\subsection{Thermalized systems ($\partial_zT\to 0$ or $|g^*|\to \infty$)}
\begin{figure}
\includegraphics[width=0.5 \columnwidth,angle=0]{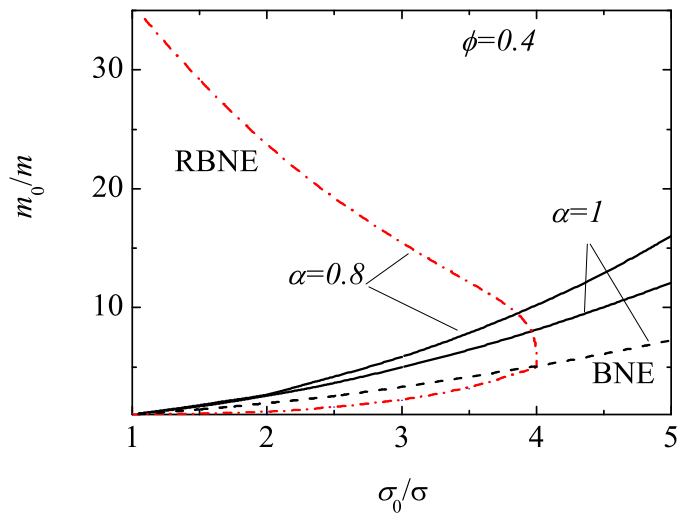}
\caption{(color online) Phase diagram for BNE/RBNE  for a two-dimensional system in the absence of thermal gradient ($|g^*|\to \infty$) at $\phi=0.4$ and two values of the (common) coefficient of restitution $\alpha=\alpha_0$. The dashed and dashed-dotted lines refer to the results obtained by Jenkins and Yoon \cite{JY02} for elastic gases ($\alpha=1$) and by Trujillo {\em et al.} \cite{TAH03} for $\alpha=0.8$, respectively.
\label{fig8}}
\end{figure}

In this Subsection the opposite limit is considered, namely, a system in which the global temperature of the bed does not vary with height ($\partial_zT\to 0$). In this case, the segregation dynamics of the intruder is only driven by the gravitational force. This is a quite interesting limit since this situation (gravity dominates the temperature gradient) can be achieved in the shaken or sheared systems employed in numerical simulations and physical experiments \cite{HQL01,WHP01,BEKR03,SBKR05}. Under these conditions ($|g^*|\to \infty$), the criterion (\ref{27}) can be written as
\begin{equation}
\label{32} \frac{1+\frac{(1+\omega)^d}{2}\chi
_{0}^{(0)}\phi(1+\alpha_{0})\frac{\gamma+M}{1+M}\frac{\Delta}{\gamma}}
{1+2^{d-2}\chi^{(0)}\phi(1+\alpha)\left[1+\phi\partial_\phi\ln
(\phi\chi^{(0)})\right]}\frac{T_0}{T}=\frac{m_0}{m}.
\end{equation}
As said in the Introduction, previous theoretical attempts to describe this particular situation have been made independently by Jenkins and Yoon \cite{JY02} for elastic systems and by Trujillo {\em et al.} \cite{TAH03} for inelastic systems. Both descriptions are based on a kinetic theory \cite{JM87} that is restricted to the quasielastic limit ($\alpha\to 1$), although Trujillo {\em et al.} \cite{TAH03} take into account the effect of non-equipartition of energy on segregation. Their segregation criterion differs from Eq.\ (\ref{32}) and is given by \cite{TAH03}
\begin{equation}
\label{33}\frac{1+\frac{(1+\omega)^d}{2}\chi
_{0}^{(0)}\phi}{1+2^{d-1}\chi^{(0)}\phi}\frac{T_0}{T}=\frac{m_0}{m},
\end{equation}
which is consistent with the one derived by Jenkins and Yoon \cite{JY02} when $\alpha=\alpha_0=1$.
The discrepancies between Eqs.\ (\ref{32}) and (\ref{33}) can be attributed to the simplicity of the kinetic theory used for deriving the latter criterion. In particular, it is easy to see that Eq.\ (\ref{32}) reduces to Eq.\ (\ref{33}) when one (i) neglects the dependence on inelasticity and assumes equipartition in certain terms, (ii) takes the approximation $\Delta =1$ (which only applies for a dilute gas of mechanically equivalent particles),
and (iii) neglects high density corrections (last term in the denominator of
(\ref{32})). Therefore, in relation to the above previous results, we can conclude that
the criterion (\ref{32}) is much more general than the one derived by Trujillo {\em et al.} \cite{TAH03} in the limit $|g^*|\to \infty$ since our results cover the complete range of the parameter space of the problem.
\begin{figure}
\includegraphics[width=0.5 \columnwidth,angle=0]{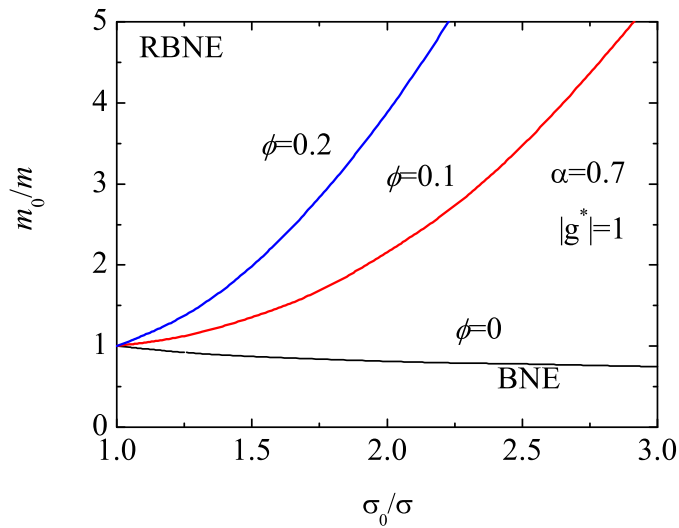}
\caption{(color online) Phase diagram for BNE/RBNE for inelastic hard spheres for $\alpha=\alpha_0=0.7$, $|g^*|=1$ and three different values of the solid volume fraction $\phi$.
\label{fig9}}
\end{figure}

A typical phase-diagram for thermalized systems delineating the regimes between BNE and RBNE is plotted in Fig.\ \ref{fig8} for the two dimensional case. (The qualitative features of the corresponding phase-diagram for the three-dimensional case are similar). Comparison between the left panel of Fig.\ \ref{fig7} (hard disks) with Fig.\ \ref{fig8} clearly shows that the presence of gravity changes dramatically the form of the phase-diagram. In particular, the main effect of inelasticity is to reduce the size of RBNE region, which is consistent with experiments \cite{BEKR03}. Moreover, we also observe that the RBNE regime appears essentially now for both large mass ratio and/or small diameter ratio. On the other hand, the predictions of Trujillo {\em et al.} \cite{TAH03} for $\alpha=0.8$ disagree with our results even at a qualitative level since they find that the mass ratio is a two-valued function of the size ratio in the phase diagram. In fact, according to the results of Trujillo {\em et al.} \cite{TAH03}, the effect of dissipation is to introduce a threshold size ratio above which there is no RBNE. We also observe that our results differ from those obtained by Jenkins and Yoon \cite{JY02} for elastic gases, especially for large size ratios.
Our results also indicate (not shown in Fig.\ref{fig8}) that non-equipartition has a weaker influence on segregation for thermalized systems than in the opposite limit ($|g^*|=0$). This behavior qualitatively agrees with the experiments carried out by Schr\"oter {\em et al.} \cite{SUKSS06} for vibrated mixtures as well as with some recent theoretical results of Yoon and Jenkins \cite{YJ06} since both works find that segregation (when is only driven by gravity) is not significantly influenced by the difference between the temperatures of the two species.

\subsection{General case}

Finally, we consider the effect of density for finite values of the reduced gravity $|g^*|$. Figure \ref{fig9} shows a phase diagram when $|g^*|=1$ (gravity comparable to the thermal gradient) for different values of the volume fraction. We have considered inlastic hard spheres ($d=3$) with $\alpha_0=\alpha=0.7$. In contrast to Fig.\ \ref{fig7bis}, we observe that the REBNE regime appears essentially now for both large mass ratio and/or small size ratio. Regarding the influence of density on the form of the phase-diagram, it is apparent that the regime of the RBNE decreases significantly with increasing $\phi$. Following Trujillo {\em et al.} \cite{TAH03}, in the fluidized regime the effect of shaking strength of vibration on the phase-diagram for BNE/RBNE can be tied to the effect of varying the solid volume fraction $\phi$. According to this argument, Fig.\ \ref{fig9} shows that the possibility of RBNE increases with increasing shaking strength (or decreasing density). The experimental findings of Breu {\em et al.} \cite{BEKR03} show similar trends with increasing shaking strength, which is consistent with our results.

\section{Conclusions}
\label{sec5}

The problem of segregation by thermal diffusion of an intruder in a dense granular gas has been addressed in this paper. Thermal diffusion is the relevant segregation mechanism in agitated granular mixtures at large shaking amplitude. In this situation, the so-called thermal diffusion factor $\Lambda$ characterizes the amount of segregation parallel to the thermal gradient \cite{KCM87}. Here, the factor $\Lambda$ has been obtained in a nonconvecting steady state with gradients only along the vertical direction (parallel to gravity). Two complementary approaches have been followed to evaluate the thermal diffusion. First, by using a hydrodynamic description $\Lambda$ has been expressed in terms of the pressure of the granular gas and the transport coefficients associated with the mass flux of impurities. Then, the above quantities have been explicitly determined by solving the inelastic Enskog equation by means of the Chapman-Enskog method \cite{CC70}. This allow us to determine $\Lambda$ as a function of the mass and size ratios, the coefficients of restitution for collisions among gas-gas and gas-intruder particles, the solid volume fraction and the reduced gravity $g^*$ [a dimensionless parameter defined in Eq.\ (\ref{11})]. Once the explicit form of $\Lambda$ is known, the condition $\Lambda=0$ provides the segregation criterion for the transition BNE$\Leftrightarrow$RBNE. This criterion is given by Eq.\ (\ref{27}) and is the most relevant result of this paper.

Although some previous theoretical efforts \cite{JY02,TAH03} on the same topic for dense granular gases have been made, they have been based on a kinetic theory which is valid for nearly elastic particles and have considered situations where gravity dominates over the temperature gradient (and so, the effects of the latter on segregation have been neglected). The present study goes beyond the weak dissipation limit and takes into account the influence of both thermal gradient and gravity (through the reduced gravity $g^*=\rho g/n\partial_zT<0$). In addition, previous results \cite{BRM05,G06} derived in the dilute regime limit are recovered at zero density ($\phi\to 0$).

To illustrate the form of the phase-diagrams BNE/RBNE in the mass and size ratio plane, two specific situations have been mainly studied: $g^*=0$ (absence of gravity) and $|g^*|\to \infty$ (homogenous temperature). The first case has been considered in recent MD simulations \cite{GDH05} while the second case has been widely studied by using kinetic theory \cite{JY02,TAH03}, computer simulations \cite{SUKSS06} and experiments \cite{BEKR03}. Our results show that the influence of dissipation on thermal diffusion is more important when the thermal gradient dominates over gravity ($g^*=0$) than in the opposite limit ($|g^*|\to \infty$). This weak influence on dissipation in the latter case contrasts with the results of Trujillo {\em et al.} \cite{TAH03} since they found the main effect of inelasticity is to introduce a threshold size ratio above which there is no RBNE (see Fig.\ \ref{fig8}). We attribute this discrepancy with Ref. \cite{TAH03} to the use of some (uncontrolled) approximations in the expressions of the partial pressures and the transport coefficients. Regarding the role played by the non-equipartition of granular energy (pseudo-thermal buoyancy force) in the segregation process, our results indicate (see Fig.\ \ref{fig7}) that while the temperature differences has an important influence on thermal diffusion in the absence of gravity, it has a weaker effect on segregation when gravity dominates over thermal gradient. These conclusions agree qualitatively well with recent MD simulations \cite{GDH05} and with some experiments carried out by Schr\"oter {\em et al.} \cite{SUKSS06} in vibrated mixtures.

Although the theory reported in this paper is consistent with previous numerical and
experimental results, a more quantitative comparison with the latter would be
desirable. As a first test, kinetic theory predictions in the Boltzmann limit ($\phi\to 0$)
\cite{G06} compare well with MD simulations of agitated dilute mixtures \cite{BRM05}.
Given that the results derived here extend the description made in Ref.\ \cite{G06} to
moderate densities, it can be reasonably expected that such a good agreement is also
kept at finite densities. In this context, it is hoped that this paper stimulates the
performance of such simulations. Another important limitation of the present study is its restriction to the tracer or intruder limit. This precludes the possibility of analyzing the influence of composition on the thermal diffusion factor. The extension of the results derived here to finite mole fraction is an interesting open problem. Moreover, this study will allow us to compare the theory with the results reported by Schr\"oter {\em et
al.} \cite{SUKSS06} in agitated mixtures constituted by particles of the same density
and equal total volumes of large and small particles. To the best of my knowledge, this
is one of the few experiments in which thermal diffusion has been isolated from the
remaining segregation mechanisms.

\acknowledgments

I am grateful to Mar\'{\i}a Jos\'e Ruiz-Montero for providing me the simulation data for Fig.\ 1. This work has been supported by the Ministerio de Educaci\'on y Ciencia (Spain) through grant No. FIS2007-60977, partially financed by
FEDER funds and by the Junta de Extremadura (Spain) through Grant No. GRU08069.

\appendix
\section{Chapman-Enskog solution in the driven case}
\label{appA}

In this Appendix we solve the Enskog-Lorentz equation (\ref{12}) to first order in the spatial gradients by means of the Chapman-Enskog method \cite{CC70}. From this solution, we determine then the transport coefficients $D_0$, $D$ and $D^T$ associated with the mass flux.

The zeroth-order distribution function $f_0^{(0)}$ obeys the Enskog equation
\begin{equation}
\label{a1}
-\frac{\zeta^{(0)} T}{2m}\frac{\partial^2}{\partial V^2}f_0^{(0)}=J_{0}^{(0)}[f_0^{(0)},f^{(0)}],
\end{equation}
where
\begin{equation}
\label{a2} J_{0}^{(0)}[f_0^{(0)},f^{(0)}]= \chi_0^{(0)}\overline{\sigma}^{d-1}\int d\mathbf{v}_{2}\int
d\widehat{{\boldsymbol\sigma}}\Theta ( \widehat{{\boldsymbol\sigma}}\cdot
\mathbf{g})(\widehat{{\boldsymbol\sigma}}
\cdot \mathbf{g})\left[ \alpha_0^{-2} f_0^{(0)}(\mathbf{v}_{1}^{\prime \prime
})f^{(0)}(\mathbf{v}_{2}^{\prime \prime})f_0^{(0)}(\mathbf{v}_{2})\right] .
\end{equation}
Upon writing Eq.\ (\ref{a1}) use has been made of the fact that $\partial_t^{(0)} T=0$ at this order in the driven case. Moreover, $\zeta^{(0)}$ refers to the cooling rate evaluated by using the zeroth-order velocity distribution function $f^{(0)}$. Since the latter is not exactly known, one has to expand $f^{(0)}$ in Sonine polynomials. A good approximation for it (at least for not very strong inelasticity) is given by the Gaussian distribution
\begin{equation}
\label{a3}
f^{(0)}(V)\to f_{M}(V)=n\left(\frac{m}{2\pi T}\right)^{d/2}e^{-mV^2/2T}.
\end{equation}
Employing it, one gets the expression (\ref{21}) for the reduced cooling rate $\zeta^*=\zeta^{(0)}/\tau$ where $\tau=n\sigma^{d-1}\sqrt{2T/m}$. The lowest order in the gradients of the cooling rate $\zeta_0^{(0)}$ for the temperature of the intruder $T_0$ can be also obtained by approximating $f_0^{(0)}$ by the Gaussian distribution
\begin{equation}
\label{a4}
f_0^{(0)}(V)\to f_{0,M}(V)=n_0\left(\frac{m_0}{2\pi T_0}\right)^{d/2}e^{-mV^2/2T_0}.
\end{equation}
With this approximation one gets expression (\ref{22}) for $\zeta_0^*=\zeta_0^{(0)}/\tau$.

The determination of the first order distribution $f_0^{(1)}$ follows similar mathematical steps as those made in the undriven case for polydisperse systems \cite{GDH07}. Here, we only display some partial results in the tracer limit ($n_0/n\to 0$). The distribution $f_0^{(1)}$ can be written as
\begin{eqnarray}
f_{0}^{(1)} &\rightarrow &\boldsymbol{\mathcal{A}}_{0}\left(
\mathbf{V}\right)\cdot  \nabla \ln
T+\boldsymbol{\mathcal{B}}_{0}\left(
\mathbf{V}\right) \cdot \nabla \ln n_{0}+\boldsymbol{\mathcal{C}}_{0}\left(
\mathbf{V}\right) \cdot \nabla \ln n \nonumber\\
&&+\mathcal{D}_{0,ij}\left( \mathbf{V} \right)
\frac{1}{2}\left( \partial _{i}U_{j}+\partial _{j
}U_{i}-\frac{2}{d}\delta _{ij}\nabla \cdot
\mathbf{U} \right)+\mathcal{E}_{0}\left( \mathbf{V} \right) \nabla \cdot
\mathbf{U}, \label{a5}
\end{eqnarray}
where the quantities $\boldsymbol{\mathcal{A}}_{0}$, $\boldsymbol{\mathcal{B}}_{0}$, $\boldsymbol{\mathcal{C}}_{0}$, $\mathcal{D}_{0,ij}$ and $\mathcal{E}_{0}$ are the solutions of a set of coupled linear integral equations.
In this paper we are only interested in the first-order contribution to the mass flux ${\bf j}_0^{(1)}$. It is defined as
\begin{equation}
{\bf j}_{0}^{(1)}=m_{0}\int d{\bf v}\,{\bf V}\,f_0^{(1)}({\bf V}). \label{a6}
\end{equation}
Use of Eq.\ (\ref{a5}) into Eq.\ (\ref{a6}) and taking into account symmetry considerations, one gets the constitutive form for ${\bf j}_0^{(1)}$ given by Eq.\ (\ref{8}) where
\begin{equation}
D^{T}=-\frac{m_0}{\rho d}\int d\mathbf{v}\mathbf{V}\cdot \boldsymbol{\mathcal{A}}_{0}\left(
\mathbf{V}\right)  \label{a7}
\end{equation}
is the thermal diffusion coefficient
\begin{equation}
D_{0}=-\frac{\rho}{m_{0}n_{0}d}\int d\mathbf{v}\mathbf{V}\cdot \boldsymbol{\mathcal{B}}_{0}\left(
\mathbf{V}\right) \label{a8}
\end{equation}
is the kinetic diffusion coefficient and
\begin{equation}
D=-\frac{1}{d}\int d\mathbf{v}\mathbf{V}\cdot \boldsymbol{\mathcal{C}}_{0}\left( \mathbf{V}\right)
\label{a9}
\end{equation}
is the mutual diffusion coefficient. According to Eqs.\ (\ref{a7})--(\ref{a9}), only the coefficients $\boldsymbol{\mathcal{A}}_{0}$, $\boldsymbol{\mathcal{B}}_{0}$, and $\boldsymbol{\mathcal{C}}_{0}$ are involved in the evaluation of the mass transport ${\bf j}_0^{(1)}$ of the intruder. These quantities are the solutions of the following set of linear integral equations:
\begin{equation}
-\frac{\zeta^{(0)} T}{2m}\frac{\partial^2}{\partial V^2}\boldsymbol{\mathcal{A}}_{0}-
J_{0}^{(0)}[\boldsymbol{\mathcal{A}}_{0},f^{(0)}]=\mathbf{A}_{0}+J_{0}^{(0)}[f_0^{(0)},\boldsymbol{\mathcal{A}}],
\label{a10}
\end{equation}
\begin{equation}
-\frac{\zeta^{(0)} T}{2m}\frac{\partial^2}{\partial V^2}\boldsymbol{\mathcal{B}}_{0}
-J_{0}^{(0)}[\boldsymbol{\mathcal{B}}_{0},f^{(0)}]=\mathbf{B}_{0}, \label{a11}
\end{equation}
\begin{equation}
-\frac{\zeta^{(0)} T}{2m}\frac{\partial^2}{\partial V^2}\boldsymbol{\mathcal{C}}_{0}
-J_{0}^{(0)}[\boldsymbol{\mathcal{C}}_{0},f^{(0)}]=\mathbf{C}_{0}+J_{0}^{(0)}[f_0^{(0)},\boldsymbol{\mathcal{C}}].
\label{a12}
\end{equation}
The inhomogeneous terms $\mathbf{A}_{0}$, $\mathbf{B}_{0}$ and $\mathbf{C}_{0}$ of the integral equations (\ref{a10})--(\ref{a12}) are defined by
\begin{equation}
A_{0,i}\left( \mathbf{V}\right)=\frac{1}{2} V_{i}\frac{\partial}{\partial {\bf V}}\cdot \left(
\mathbf{V}f_{0}^{(0)}\right) -\frac{p}{\rho}\frac{\partial f_{0}^{(0)}}{\partial V_{i}}
+\frac{1}{2}\mathcal{ K}_{0,i}\left[\frac{\partial}{\partial {\bf V}}\cdot \left( \mathbf{V}
f^{(0)}\right) \right] , \label{a13}
\end{equation}
\begin{equation}
{\bf B}_{0}\left( \mathbf{V}\right) = -{\bf V} f_{0}^{(0)},  \label{a14}
\end{equation}
\begin{equation}
\label{a15} C_{0,i}\left( \mathbf{V}\right) =-m^{-1}\frac{\partial p}{\partial n}
\frac{\partial f_{0}^{(0)}}{\partial V_{i}}
- \frac{(1+\omega)^{-d}}{\chi_{0}^{(0)}T}\left(\frac{\partial \mu_0}{\partial \phi}\right)_{T,n_0}\mathcal{K}_{0,i}\left[f^{(0)}\right].
\end{equation}
In Eqs.\ (\ref{a13})--(\ref{a15}), the pressure $p$ is given by \cite{GD99,L05}
\begin{equation}
p=nT\left[1+2^{d-2}\chi \phi(1+\alpha)\right], \label{a15.1}
\end{equation}
$\sigma\equiv \sigma_0/\sigma$ is the size ratio, $\mu_0$ is the chemical potential of the intruder and the operator $\mathcal{K}_{0,i}[X]$ is defined as
\begin{equation}
\label{a18}
\mathcal{K}_{0,i}[X] =\overline{\sigma}^{d}\chi _{0}^{(0)}\int d \mathbf{v}_{2}\int d\widehat{\boldsymbol
{\sigma }}\Theta (\widehat{\boldsymbol {\sigma}} \cdot \mathbf{g})(\widehat{\boldsymbol {\sigma}}\cdot
\mathbf{g})
\widehat{\sigma}_i
\left[ \alpha _{0}^{-2}f_{0}^{(0)}(\mathbf{V}_{1}^{\prime \prime})X(\mathbf{V}_{2}^{\prime \prime
})+f_{0}^{(0)}(\mathbf{V}_{1})X(\mathbf{V}_{2})\right] .
\end{equation}
Note that, in contrast to what happens in the undriven case \cite{GDH07}, here each one of the quantities $\boldsymbol{\mathcal{A}}_{0}$, $\boldsymbol{\mathcal{B}}_{0}$, and $\boldsymbol{\mathcal{C}}_{0}$ obey closed integral equations. Moreover, upon writing Eqs.\ (\ref{a10})--(\ref{a12}), use has been made of the expression of the first-order distribution $f^{(1)}$ of the gas particles. Its form in the driven heated case has been derived in Ref.\ \cite{GM02} and reads
\begin{eqnarray}
f^{(1)} &\rightarrow &\boldsymbol{\mathcal{A}}\left(
\mathbf{V}\right)\cdot  \nabla \ln
T+\boldsymbol{\mathcal{C}}\left(
\mathbf{V}\right) \cdot \nabla \ln n\nonumber\\
&&+\mathcal{D}_{ij}\left( \mathbf{V} \right)
\frac{1}{2}\left( \partial _{i}U_{j}+\partial _{j
}U_{i}-\frac{2}{d}\delta _{ij}\nabla \cdot
\mathbf{U} \right)+\mathcal{E}\left( \mathbf{V} \right) \nabla \cdot
\mathbf{U}. \label{a18.1}
\end{eqnarray}

For practical purposes, the integral equations (\ref{a10})--(\ref{a12}) must be approximately solved by using a Sonine polynomial expansion. In the lowest Sonine approximation, the quantities $\boldsymbol{\mathcal{A}}_0$, $\boldsymbol{\mathcal{B}}_0$ and
$\boldsymbol{\mathcal{C}}_0$ are approximated by
\begin{equation}
\label{a19} \boldsymbol{\mathcal{A}}_{0}({\bf V})\to -f_{0,M}({\bf V})\frac{\rho}{n_0T_0}{\bf
V}D^T,
\end{equation}
\begin{equation}
\label{a20} \boldsymbol{\mathcal{B}}_{0}({\bf V})\to -f_{0,M}({\bf V})\frac{m_0^2}{\rho T_0}{\bf
V}D_{0},
\end{equation}
\begin{equation}
\label{a21} \boldsymbol{\mathcal{C}}_{0}({\bf V})\to -f_{0,M}({\bf V})\frac{m_0}{ n_0T_0}{\bf
V}D,
\end{equation}
where $f_{0,M}$ is given by Eq.\ (\ref{a4}). Consistently, $\boldsymbol{\mathcal{A}}$ and  $\boldsymbol{\mathcal{C}}$ must be also approximated in a similar way. However, both quantities vanish in the lowest Sonine approximation \cite{GD99b}. To get the transport coefficients $D^T$, $D_0$ and $D$, we substitute first $\boldsymbol{\mathcal{A}}_0$, $\boldsymbol{\mathcal{B}}_0$ and $\boldsymbol{\mathcal{C}}_0$ by their Sonine approximations (\ref{a19})--(\ref{a21}), respectively, and then, we multiply the integral equations (\ref{a10})--(\ref{a12}) by $m_0{\bf V}$ and integrate over the velocity. After some algebra, one gets
\begin{equation}
\label{a22} \nu_{D} D^T=-\frac{x_0pm_0}{m\rho}\left(1-\frac{\rho
T_0}{m_0p}\right)- \frac{1}{2d\rho}\int d{\bf V}\; m_0 V_i \mathcal{ K}_{0,i
}\left[\frac{\partial}{\partial {\bf V}}\cdot \left( \mathbf{V} f^{(0)}\right) \right],
\end{equation}
\begin{equation}
\label{a23} \nu_{D} D_0=\frac{\rho T_0}{m_0^2},
\end{equation}
\begin{equation}
\label{a24} \nu_{D} D=-\frac{x_0 M n}{m_0}\frac{\partial p}{\partial n}+ \frac{(1+\omega)^{-d}}{d\chi_{0}^{(0)}T}\left(\frac{\partial \mu_0}{\partial \phi}\right)_{T,n_0}\int d{\bf V}\; V_i \mathcal{ K}_{0,i}\left[f^{(0)}\right],
\end{equation}
where $x_0=n_0/n$, $M\equiv m_0/m$ is the mass ratio and
\begin{equation}
\label{a25} \nu_{D}=-\frac{1}{dn_0T_0}\int d{\bf v}\;m_0{\bf V}\cdot
J_{0}^{(0)}[f_{0,M}{\bf V},f^{(0)}].
\end{equation}
Note that Eqs.\ (\ref{a22})--(\ref{a24}) have been obtained by neglecting some non-Gaussian contributions to the zeroth-order distribution $f_0^{(0)}$. The collision integrals appearing on the right hand side of Eqs.\ (\ref{a22}) and (\ref{a24}) along with the collision frequency $\nu_D$ have been evaluated in Ref.\ \cite{GHD07} in the case of a multicomponent mixture. In the tracer limit, one easily gets
\begin{equation}
\label{a26} \frac{1}{2d\rho}\int\;  d{\bf V}\; m_0 V_i \mathcal{ K}_{0,i
}\left[\frac{\partial}{\partial {\bf V}}\cdot \left( \mathbf{V} f^{(0)}\right)
\right]=-\frac{1}{2}\frac{x_0T}{m}(1+\omega)^d \frac{M}{1+M}\phi
 \chi_0^{(0)} (1+\alpha_0),
\end{equation}
\begin{equation}
\label{a27}
\frac{1}{d}\int d{\bf V}\; m_0 V_i \mathcal{ K}_{0,i}\left[f^{(0)}\right]=\frac{1}{2}x_0 n T(1+\omega)^d \frac{\gamma+M}{1+M}\phi\chi_0^{(0)} (1+\alpha_0),
\end{equation}
\begin{equation}
\label{a28} \nu_D=\frac{2\pi^{(d-1)/2}}{d\Gamma\left(\frac{d}{2}\right)}n\overline{\sigma}^{d-1}\sqrt{\frac{2T}{m}}
\frac{\chi_0^{(0)}}{1+M}\sqrt{\frac{\gamma+M}{M}}(1+\alpha_0),
\end{equation}
where $\gamma=T_0/T$ is the temperature ratio. The final expressions for the reduced transport coefficients defined by (\ref{10}) can be easily obtained from Eqs.\ (\ref{a22})--(\ref{a24}) and (\ref{a26})--(\ref{a28}). They are given by Eqs.\ (\ref{16})--(\ref{18}) with $\nu_D^*\equiv \nu_D/\tau$.


\begin{thebibliography} {99}
\bibitem{note}Note that the form of the effective collision frequency $\tau$ chosen here differs sligthly from the one considered in Ref.\ \cite{G08}.
\end{thebibliography}
\end{document}